\documentstyle[12pt]{article}

\begin{document}

\renewcommand{\thefootnote}{\fnsymbol{footnote}}

\thispagestyle{empty}

\hfill \parbox{45mm}{
{MPI-PhT/95-44}
\par
May 1995 \par
hep-th/9505094}

\vspace*{15mm}

\begin{center}
{\LARGE Theoretical Analysis of a Reported}

\smallskip

{\LARGE Weak Gravitational Shielding Effect.}
        
\vspace{22mm}

{\large Giovanni Modanese}%
\footnote{A. Von Humboldt Fellow.\par
\ \ \ e-mail: modanese@science.unitn.it}

\medskip

{\em Max-Planck-Institut f\"ur Physik \par
Werner-Heisenberg-Institut \par
F\"ohringer Ring 6, D 80805 M\"unchen (Germany)}

\bigskip \bigskip

{To appear in: Europhys.\ Lett.}

\medskip

\end{center}

\vspace*{10mm}

\begin{abstract}
Under special conditions (Meissner-effect levitation in a high frequency 
magnetic field and rapid rotation) a disk of high-$T_c$ superconducting 
material has recently been found to produce a weak shielding of the 
gravitational field. We show that this phenomenon has no explanation in 
the standard gravity theories, except possibly in the non-perturbative 
Euclidean quantum theory. 

\medskip
\noindent
04.20.-q Classical general relativity.

\noindent
04.60.-m Quantum gravity.

\noindent
74.72.-h High-$T_c$ cuprates.

\bigskip 

\end{abstract}

In two recent experiments \cite{p1,p2}, Podkletnov and co-workers 
have found indications for a possible weak shielding of the 
gravitational force through a disk of high-$T_c$ superconducting 
material. In the first experiment a sample made of silicon
dioxide of the weight of ca.\ 5 $g$, was found to lose about
0.05\% of its weight when placed 15 $mm$ above the disk.
The diameter of the disk was 145 $mm$ and its thickness 
6 $mm$. The disk was refrigerated using liquid helium and was 
levitating over a solenoid due to the Meissner effect. When the 
disk was set in rotation by means of lateral alternating magnetic 
fields, the shielding effect increased up to 0.3\%. 
When the disk was not levitating, but was placed over a fixed 
support, no shielding was observed.

In the second experiment the disk had the form of a toroid
with the outer diameter of 275 $mm$ and was enclosed in a stainless
steel cryostat. Samples of different composition and weight 
(10 to 50 $g$) were placed over the disk and the same percentual
weight loss was observed for different samples, thus enforcing
the interpretation of the effect as a slight diminution of the 
gravitational acceleration. While the toroid was rotating (at an
angular speed of 5000 rpm) the weight loss was of 0.3-0.5\%, like in 
the first experiment, but it reached a maximum of 1.9-2.1\% when the 
speed was slowly reduced by varying the current in the solenoids.

In both experiments, the magnetic fields were produced by high 
frequency currents and the maximum effect was observed at frequencies 
of the order of ca.\ 1 $MHz$. Measurements were effected also
in the vacuum, in order to rule out possible buoyancy effects.
The dependence of the shielding
value on the height above the disk was very weak. Within the 
considered range (from a few $cm$ to 300 $cm$) no sensible variation 
of the shielding value was observed. This weak height dependence is
a severe challenge for any candidate theoretical interpretation,
as it violates an intuitive vectorial representation of the
shielding. We have analyzed in detail this issue in \cite{up}.

Independent repetitions of the experiment have already been 
undertaken, stimulated by scientific and especially technological
interest. We would like to stress here the importance of precise
measurements. In particular, it is essential to obtain
exact spatial field maps and information about the
transient stages. It is also crucial to use a different kind
of balance from that used by the authors of \cite{p1,p2}
and possibly a gravity gradiometer \cite{paik}. If the effect turned out
to be of non-gravitational nature, its fundamental interest
would be strongly reduced. On the contrary, if the effect is
really a gravitational shielding its theoretical explanation
calls for new and non-trivial dynamical mechanisms, as
we argue in the following.

For clarity we shall organize our analysis as follows. Considering
two masses $m_1$ and $m_2$ (which represent the Earth and the 
sample) and a medium between them (the disk), we shall evaluate
their potential energy in these alternatives:

\medskip \noindent
(1) the {\it medium} can be regarded as a classical system, as a quantum
system, or as a Bose condensate with macroscopic wave function;

\medskip \noindent
(2) the {\it gravitational field} can be regarded as classical or
quantized, and in both cases as weak (perturbations theory) or strong.

\medskip
In which of these approximations could a shielding effect arise?
We can immediately observe that the possibilities of Point (1) are
severely restricted by a large body of experimental evidence. 
Namely a sensible gravitational shielding has never previously been 
observed. Several experiments, starting with the classical
measurements of Q.\ Majorana, have shown that the gravitational force 
is not influenced by any medium, up to one part in $10^{10}$ or less
(for a very complete list of references see \cite{gillies}).
For a ``classical'' medium the reason for this is essentially the 
absence of charges of opposite sign which, by shifting or migrating 
inside the medium, might generate a field which counteracts the 
applied field. On the other hand when the medium is regarded as
a quantum system, it is easy to check that the probability 
of a (virtual) process in which a graviton excites an atom or a 
molecule of the medium and is absorbed is exceedingly small, essentially 
due to the smallness of the gravitational coupling at the atomic 
level (see for instance \cite{weinberg}). It is then clear that 
the reported shielding can only be due to the Bose condensate present 
in the high-$T_c$ superconductor. 

Coming to Point (2), first we regard the gravitational 
field as classical. It is readily realized that in general neither 
the superconducting disk nor any other object of reasonable 
density, if placed close to the sample mass, can influence 
the local geometry so much as to modify its weight by the observed 
amount. To check this one just needs to write the Einstein
equations (or even some generalizations of them) and impose suitable
conditions on the source $T_{\mu \nu}$:
\begin{eqnarray}
  & & R_{\mu \nu} - \frac{1}{2} g_{\mu \nu} R = -8\pi G T_{\mu \nu}; \\
  & & G \sim 10^{-66} \ cm^2 \ \ \ {\rm in \ natural \ units};
  \nonumber \\
  & & |T_{\mu \nu}| < ... \nonumber
\label{einstein}
\end{eqnarray}
Let us be more explicit: according to Einstein equations any
apparatus with mass-energy comparable to that of Ref.s \cite{p1,p2},
if placed far away from any other source of gravitation, is unable 
to produce a gravitational field of the intensity of ca.\ 0.01 $g$
\footnote{When examining this possibility one should hypothesize that
the disk produces a repulsive force. But it is known
that arguments in favour of "antigravity" are untenable
\cite{ag} and that local negative energy densities are strongly
constrained in Quantum Field Theory (see for instance \cite{ford}). 
There remains only the possibility of gravitomagnetic and 
gravitoelectric effects, which are however usually very small 
\cite{jap}. In \cite{li} it is argued using the Maxwell-like
approximated form of Einstein equations that the gravitoelectric
field produced by a superconductor could be abnormally strong.
In our opinion this conclusion contrasts with the full equations
(\ref{einstein}). For a comparison, consider the strength of the
"gravitational Meissner effect": in a neutron star with density
of about $10^{17} \ kg/m^3$ the gravitational London penetration
depth is ca.\ 12 $km$ \cite{lano}. An experimental check disproving 
the hypothesis of a repulsive force is the measurement of $g$ 
{\it below} the disk. Preliminary measurements \cite{pk3} do not 
show any variation of $g$.}.
The shielding effect, if true, must then consist of some kind of
"absorbtion" of the Earth's field in the superconducting disk.

Having thus excluded any possibility of shielding for a classical
gravitational field, we need now an expression for the gravitational
potential energy of two masses $m_1$ and $m_2$ which takes into 
account quantum field effects, possibly also at non-perturbative level.
This is given in Euclidean quantum gravity by the functional integral
\cite{m1}
\begin{eqnarray}
  E & = & \lim_{T \to \infty} - \frac{\hbar}{T}
  \log \frac{\int d[g] \, \exp \left\{ - \hbar^{-1} \left[
  S_g + \sum_{i=1,2} m_i \int_{-\frac{T}{2}}^{\frac{T}{2}} dt \,
  \sqrt{g_{\mu \nu}[x_i(t)] \dot{x}_i^\mu(t) \dot{x}_i^\nu(t)}
  \right] \right\}}{\int d[g] \, \exp \left\{ - \hbar^{-1} S_g \right\} }
  \label{ciao} \\
  & \equiv & \lim_{T \to \infty} - \frac{\hbar}{T} \log 
  \left< \exp \left\{ - \hbar^{-1} \sum_{i=1,2} m_i 
  \int_{-\frac{T}{2}}^{\frac{T}{2}} ds_i \right\} \right>_{S_g} .
\label{bella}
\end{eqnarray}
where $S_g$ is the gravitational action
\begin{equation}
  S_g = \int d^4x \, \sqrt{g} \left( \frac{\Lambda}{8\pi G} - 
  \frac{R}{8\pi G} + \frac{1}{4} a R_{\mu \nu \rho \sigma} 
  R^{\mu \nu \rho \sigma} \right) .
\label{azione}
\end{equation}

The $R^2$ term in $S_g$ (important only at very small scale) is necessary 
to ensure the positivity of the action. The trajectories $x_i(t)$ of 
$m_1$ and $m_2$ are parallel with respect to the metric $g$; let $L$ be 
the distance between them, corresponding to the spatial distance of the two
masses. An evaluation of (\ref{bella}) in the non-perturbative lattice 
theory has ben carried out recently by Hamber and Williams \cite{hw}.

In perturbation theory \cite{veltman} the metric 
$g_{\mu \nu}(x)$ is expanded in the traditional way as the sum of 
a flat background $\delta_{\mu \nu}$ plus small fluctuations 
$\kappa h_{\mu \nu}(x)$ ($\kappa=\sqrt{8\pi G}$). The cosmological 
and the $R^2$ terms are dropped, leaving the pure Einstein action. 
Eq.\ (\ref{ciao}) is rewritten as
\begin{equation}
  E = \lim_{T \to \infty} - \frac{\hbar}{T} 
  \log \frac{\int d[h] \, \exp \left\{ - \hbar^{-1} \left[
  S_{\rm Einst.} + \sum_{i=1,2} m_i
  \int_{-\frac{T}{2}}^{\frac{T}{2}} dt \, 
  \sqrt{1 + h_{00}[x_i(t)] } \right] \right\} }
  {\int d[h] \, \exp \left\{ - \hbar^{-1} S_{\rm Einst.} \right\} } ,
\label{gioia}
\end{equation}
where the trajectories $x_1(t)$ and $x_2(t)$ are two parallel lines
in flat space. Expanding (\ref{gioia} ) in powers of $\kappa$ one 
obtains to lowest order the Newton potential \cite{m1}, and to 
higher orders its relativistic and quantum corrections \cite{mv}. 

The Bose condensate composed by the Cooper pairs inside the superconductor 
is described by a bosonic field $\phi$ with non-vanishing 
vacuum expectation value $\phi_0=\langle 0 | \phi | 0 \rangle$. 
Using the notation $\phi = \phi_0 + \tilde{\phi}$, the action of
such a field coupled to the gravitational field has the form
\begin{eqnarray}
  & S_\phi = \int d^4x \, \sqrt{g(x)} & \Bigl\{
  \partial_\mu \left[ \phi_0(x) + \tilde{\phi}(x) \right]^*
  \partial_\nu \left[ \phi_0(x) + \tilde{\phi}(x) \right]
  g^{\mu \nu}(x) \nonumber \\
  & & + \frac{1}{2} m^2 |\phi_0(x)|^2 + \frac{1}{2} m^2
  \left[ \phi_0^*(x) \tilde{\phi}(x) + \phi_0(x) \tilde{\phi}^*(x) \right]
  + \frac{1}{2} m^2 |\tilde{\phi}(x)|^2 \Bigr\} ,
\end{eqnarray}
where $m$ is the mass of a Cooper pair. In order to describe the interaction
we insert $S_\phi$ into (\ref{ciao}) and include $\tilde{\phi}$ into
the integration variables, while $\phi_0$ is considered as an external 
source, being determined essentially by the structure of the
superconductor and by the external e.m.\ fields. In the following we 
shall disregard in $S_\phi$ the terms containing $\tilde{\phi}$, as they 
give rise to emission-absorption processes of gravitons which we know 
to be irrelevant.

Perturbatively, the interaction of $h_{\mu \nu}(x)$ with the condensate 
$\phi_0(x)$ is principally mediated by the vertex
\begin{equation}
  {\cal{L} }_{h\phi_0} = \kappa \partial_\mu \phi_0^*(x) \partial_\nu 
  \phi_0(x) h^{\mu \nu}(x).
\label{vertice}
\end{equation}
This produces corrections to the gravitational propagator, which are
however practically irrelevant, because they are proportional to 
powers of $\kappa \sim 10^{-33} \ cm$. It is straightforward to 
compute the corresponding corrections to eq.\ (\ref{gioia}).
We do not need to investigate in detail the signs of these
corrections or their dependence from $\phi_0$: they are in any
case too small (by several magnitude orders) to account for the 
reported shielding effect.

Looking at the total action $S=S_g+S_\phi$ we recognize besides
the familiar vertex (\ref{vertice}) a further coupling between
$g_{\mu \nu}(x)$ and $\phi_0(x)$. Namely, the Bose condensate
contributes to the cosmological term. We can rewrite the total
action (without the $R^2$ term for gravity) as
\begin{equation}
  S = S_g + S_\phi = \int d^4 x \, \sqrt{g(x)}
  \left\{ \left[ \frac{\Lambda}{8\pi G} + \frac{1}{2} \mu^2(x)
  \right] - \frac{R}{8\pi G} \right\} + S_{h \phi_0}
  + S_{\tilde{\phi}},
\label{s}
\end{equation}
where
\begin{eqnarray}
  \frac{1}{2} \mu^2(x) & = & \frac{1}{2} \left[ \partial_\mu \phi_0(x) 
  \right]^* \left[ \partial^\mu \phi_0(x) \right]
  + \frac{1}{2} m^2 |\phi_0(x)|^2 , \\
  S_{h \phi_0} & = & \int d^4 x \, \sqrt{g(x) } {\cal{L} }_{h \phi_0}
\end{eqnarray}
and $S_{\tilde{\phi}}$ comprises the terms which contain at least
one field $\tilde{\phi}$ and are thus irrelevant, as we mentioned
above.

We see from (\ref{s}) that the condensate $\phi_0(x)$ and its
four-dimensional gradient give a positive contribution to the 
intrinsic cosmological term $\Lambda/8\pi G$. It is known that 
a positive cosmological constant turns Eintein gravity into an 
unstable theory, as it corresponds in the action to a mass term 
with negative sign \cite{veltman}. In the case we are considering 
here, the cosmological term is spacetime dependent. The situation 
is thus quite complicated and we shall just sketch it briefly in 
the following; a more complete account can be found in \cite{cc}.

Since in any four-dimensional region $\Omega$ in which $\mu^2 > 
|\Lambda|/8\pi G$ the mass term of the gravitational field 
is non-zero and negative, in this region the field
can grow without limit, at least classically (suppose to
minimize the action (\ref{azione}) by trial functions which
are vacuum solutions ($R=0$), disregarding the $R^2$ term).
In fact there will be some physical cut-off; thus within $\Omega$
the gravitational field will be forced to some fixed value,
independent of the external conditions. This is similar to what 
happens in electrostatics in the presence of perfect conductors: 
the electric field is constrained to be zero within the conductors.
In that case the physical origin of the constraint is different
(a redistribution of opposite charges); but in both cases the effect 
on the field propagator turns out to be that of a shielding.

At this point we would need an estimate for both $|\Lambda|/8\pi G$ and 
$\mu^2$. It is known that the cosmological constant observed at
astronomical scales is very small; a typical upper limit is
$|\Lambda|G < 10^{-120}$, which means $|\Lambda|/8\pi G<
10^{12} \ cm^{-4}$. To estimate $\mu^2$ we can assume
an average density of Cooper pairs in the superconductor of
$\sim 10^{20} \ cm^{-3}$. Remembering that the mass of
a pair is $\sim 10^{10} \ cm^{-1}$ in natural units,
we find that $\phi_0 \sim 10^5 \ cm^{-1}$ and thus
$m^2 |\phi_0|^2 \sim 10^{30} \ cm^{-4}$. If $\phi_0$
varies over distances of the order of $10^{-8} \div 10^{-7} \ cm$, 
as usual in high-$T_c$ superconductors, the gradient $(\partial \phi_0 / 
\partial x)$ can be of the order of $10^{12} \ cm^{-2}$. 

These values support our hypothesis that the total cosmological
term is positive in the superconductor. Actually the positive
contribution of the condensate is such that one could expect
the formation of gravitational singularities in any superconductor,
subjected to external fields or not -- a fact which contrasts
with the observations. This can be avoided if we take the point
of view, supported by lattice quantum gravity with a fundamental
length \cite{hw,cc,m2}, that the effective intrinsic cosmological constant
depends on the momentum scale $p$ like
\begin{equation}
  |\Lambda|G \sim (\ell_0 p)^\gamma ,
\label{scala}
\end{equation}
where $\ell_0 \sim \kappa$ is the fundamental lattice length
(Planck length) and $\gamma$ is a critical exponent which up to
now has been computed only for small lattices. The sign of
$\Lambda$ is negative. This provides a well defined flat space
limit for the non-perturbative Euclidean theory based on the
action (\ref{azione}), and ensures that the signs in our
discussion of the minimization of the total action are correct.

We hypothesize that at the length scale of $10^{-8} \div 10^{-6}
\ cm$ $\Lambda$ could be of the same order of the average of
$\mu^2$, so that the competition between the two terms would
lead to singularities only in those regions of the superconductor
where the condensate density is larger than elsewhere. From this
hypothesis we deduce that $|\Lambda| \sim 10^{-36} \ cm^{-2}$, a value
well compatible with the conventional experimental data at that scale
\footnote{It corresponds to a graviton mass of
$\sim 10^{-18} \ cm^{-1}$ and thus to a (unobservable) range of gravity 
of $\sim 10^{18} \ cm$. Note that a similar argument allows to restrict
the exponent $\gamma$ in eq.\ (\ref{scala}) to $\gamma>2$.
In fact, for $\gamma=2$ we would have $|\Lambda| \sim p^2$, in
contradiction with the fact that in order to be unobservable the 
range of gravity must be much larger than the length scale 
$p^{-1}$ which we are considering.}.

In conclusion, a shielding effect of the reported magnitude cannot
be explained by classical General Relativity, nor by the usual
perturbation theory of quantum gravity coupled to the Cooper-pair
density $\phi_0(x)$ through the vertex (\ref{vertice}).
We have then considered a further possible coupling mechanism:
the term $\mu^2(x) = [\partial^\mu \phi_0^*(x)]
[\partial_\nu \phi_0(x)] + m^2 \phi_0^2(x)$ in the condensate's
action may act as a positive contribution to the effective gravitational
cosmological constant. This may produce localized gravitational
instabilities and thus an observable effect, in spite of the
smallness of $G$, which makes the coupling (\ref{vertice}) very
weak. In the regions where the gravitational field becomes unstable
it tends to take "fixed values" independent of
the neighboring values. This reminds of the situation of electrostatics
in the presence of perfect conductors. The effect on the field
propagator and on the static potential is that of a partial shielding.

According to this picture, the magnetic fields which keep the
disk levitating and rotating and produce currents inside it
play the role (together with the microscopic structure of the HTC 
material) of determining the condensate density $\phi_0(x)$. The
energy necessary for the partial "absorbtion" of the
gravitational field is supplied by the high-frequency components
of the magnetic field.

The autor would like to thank D.\ Maison and the staff of MPI for the 
kind hospitality in Munich, and the A.\ von Humboldt Foundation for
financial support. Useful suggestions of G.\ Fiore, G.\ Fontana,
F.\ Pavese, V.\ Polushkin, D.\ Saam and R.\ Stirniman are acknowledged
too.

\end{document}